\documentclass[aps,prd,twocolumn,superscriptaddress,floatfix]{revtex4-1}
\usepackage{graphicx}
\usepackage{bm}
\usepackage{amsmath}
\usepackage{amssymb}
\usepackage{amsthm}

\newcommand{\be}{\begin{equation}}
\newcommand{\ee}{\end{equation}}
\newcommand{\bea}{\begin{eqnarray}}
\newcommand{\eea}{\end{eqnarray}}
\newcommand{\bdm}{\begin{displaymath}}
\newcommand{\edm}{\end{displaymath}}
\newcommand{\beas}{\begin{eqnarray*}}
\newcommand{\eeas}{\end{eqnarray*}}

\begin{document}
\title{Fundamental Cosmology from Precision Spectroscopy: II. Synergies with supernovae}
\author{A. C. O. Leite}  
\email[]{Ana.Leite@astro.up.pt}
\affiliation{Centro de Astrof\'{\i}sica da Universidade do Porto, Rua das Estrelas, 4150-762 Porto, Portugal}
\affiliation{Faculdade de Ci\^encias, Universidade do Porto, Rua do Campo Alegre, 4150-007 Porto, Portugal}
\author{C. J. A. P. Martins}
\email[]{Carlos.Martins@astro.up.pt}
\affiliation{Centro de Astrof\'{\i}sica da Universidade do Porto, Rua das Estrelas, 4150-762 Porto, Portugal}
\affiliation{Instituto de Astrof\'{\i}sica e Ci\^encias do Espa\c co, CAUP, Rua das Estrelas, 4150-762 Porto, Portugal}
\date{23 March 2015}

\begin{abstract}
In previous work [Amendola {\it et al.}, Phys. Rev. D86 (2012) 063515], Principal Component Analysis based methods to constrain the dark energy equation of state using Type Ia supernovae and other low redshift probes were extended to spectroscopic tests of the stability fundamental couplings, which can probe higher redshifts. Here we use them to quantify the gains in sensitivity obtained by combining spectroscopic measurements expected from ESPRESSO at the VLT and the high-resolution ultra-stable spectrograph for the E-ELT (known as ELT-HIRES) with future supernova surveys. In addition to simulated low and intermediate redshift supernova surveys, we assess the dark energy impact of high-redshift supernovas detected by JWST and characterized by the E-ELT or TMT. Our results show that a detailed characterization of the dark energy properties beyond the acceleration phase (i.e., deep in the matter era) is viable, and may reach as deep as redshift 4. 
\end{abstract}

\keywords{Cosmology: Theory, Dark Energy, Variation of alpha}
\pacs{98.80.-k,98.80.Jk}
\maketitle

\section{\label{intro}Introduction}

The observational evidence for the acceleration of the universe demonstrates that our canonical theories of cosmology and fundamental physics are at least incomplete, and possibly incorrect. Substantial efforts are being put, therefore, into identifying the mechanisms responsible for it---see \cite{Weinberg} for a recent review. At the most basic level one would like to know if the acceleration is due to a cosmological constant (the simplest viable possibility), to a modification of the behavior of gravity on large scales, or to a cosmological dynamical scalar field, but ultimately the goal must be to characterize its behavior as a function of redshift. Mapping the redshift dependence of the dark energy equation of state is a simple, standard way to do this.

Astrophysical tests of the stability of fundamental couplings, such as the fine-structure constant $\alpha$, are among the most powerful probes of the equation of state of dynamical dark energy \cite{Wetterich,Carroll}. The idea was qualitatively discussed in \cite{Dvali,Chiba,Nunes,Doran,Avelino}, and more recently forecasts of its impact have been obtained \cite{Amendola,Leite}. This complements other methods due to its large redshift lever arm and the fact that these measurements can be done from ground-based facilities, both in the UV/optical and the radio/mm bands. While currently available measurements have moderate sensitivity \cite{Dipole} and control of systematics is a possible concern---see \cite{Uzan} for a recent overview of the theoretical context and observational status---this sensitivity is improving \cite{LP1,LP3}, and the prospects are much better for the forthcoming generation of observational facilities \cite{GRG}.

In \cite{Amendola} we extended previously available Principal Component Analysis (PCA, see e.g. \cite{Huterer}) methods that had been used in other contexts in cosmology to the case of tests of the stability of fundamental couplings, and these were then calibrated with some current data from the UVES spectrograph in \cite{Leite}. Here we further quantify the potential of this method, by discussing the improvements in dark energy characterization that can be obtained by combining these spectroscopic measurements with those of future supernova surveys at various redshifts. We consider two different fiducial models and several $\alpha$ and supernova samples (with different sensitivities and redshift ranges) in order to quantify the effect of these parameters. In the next section we review the relevant PCA methodology, and in Sect. III we spell out our assumptions on models and data. Sect. IV contains our main results, quantitatively expressed in terms of a figure of merit, but we also provide a visual illustration by showing examples of dark energy equation of state reconstructions. Finally we present our conclusions in Sect. V.


\section{\label{methods}PCA tools}

Our study is based on non-parametric PCA techniques described in detail in \cite{Huterer,Amendola}. Here we provide a short summary of the key features for our analysis. We divide the relevant redshift range into $N$ bins, with the equation of state taking the value $w_i$ at bin $i$,
\begin{equation}
w(z) = \sum_{i = 1}^N w_i \theta_i(z) \,.
\end{equation}
Another way of saying this is that $w(z)$ is expanded in the basis $\theta_i$, with $\theta_1 =(1,0,0,...)$, $\theta_2 = (0,1,0,...)$, etc. The precision on the measurement of $w_i$ can be inferred from the Fisher matrix of the parameters $w_i$, specifically from $\sqrt{(F^{-1})_{ii}}$. One can find a basis in which all the parameters are uncorrelated by diagonalizing the Fisher matrix such that $F = W^T \Lambda W$, where $\Lambda$ is diagonal and the rows of $W$ are the eigenvectors $e_i(z)$ or the principal components. These define the new basis in which the new coefficients $\alpha_i$ are uncorrelated, and we can write
\begin{equation}
\label{recw} w(z) = \sum_{i = 1}^N \alpha_i e_i(z) \,.
\end{equation}
The diagonal elements of $\Lambda$ are the eigenvalues $\lambda_i$ (ordered from largest to smallest) and define the variance of the new parameters, $\sigma^2(\alpha_i) = 1/\lambda_i$.

We will consider quintessence-type models where the same scalar field yields the dark energy and (through a coupling to the electromagnetic sector) a variation of the fine-structure constant $\alpha$ \cite{Carroll,Dvali,Chiba,GRG}. We take this coupling to be
\be
{\cal L}_{\phi F} = - \frac{1}{4} B_F(\phi) F_{\mu\nu}F^{\mu\nu} ,
\ee
where the gauge kinetic function $B_F(\phi)$ is linear,
\be
B_F(\phi) = 1- \zeta \kappa (\phi-\phi_0) ,\label{coupling}
\ee
$\kappa^2=8\pi G$ and $\zeta$ is the coupling constant, which in what follows will be marginalized over. This can be seen as the first term of a Taylor expansion, and should be a good approximation if the field is slowly varying at low redshift. Then the evolution of $\alpha$ is given by
\begin{equation}
\frac{\Delta \alpha}{\alpha}(z) \equiv
\frac{\alpha(z)-\alpha_0}{\alpha_0} = \zeta \kappa (\phi(z)-\phi_0) \,.
\end{equation}
For a flat Friedmann-Robertson-Walker Universe with a canonical scalar field, $\dot{\phi}^2 = [1+w(z)]\rho_\phi$, and therefore, for a given dependence of the equation of state parameter $w(z)$ with redshift, the scalar field evolution can be written
\begin{equation}
\phi(z)-\phi_0 = \frac{\sqrt{3}}{\kappa}
\int_0^z \sqrt{1+w(z)} \left(1+ \frac{\rho_m}{\rho_\phi}\right)^{-1/2}
\frac{dz}{1+z} .
\end{equation}
where we have chosen the positive root of the solution, and therefore the relative variation of $\alpha$ is given by
\begin{equation}
\frac{\Delta\alpha}{\alpha}(z)=\zeta \int_0^z \sqrt{3[1+w(z)]\Omega_\phi(z)}\frac{dz}{1+z}\,,
\end{equation}
where $\Omega_\phi=\rho_\phi/(\rho_m+\rho_\phi)$ is the fraction of the universe's energy in the scalar field.

From the above one can calculate the Fisher matrix using standard techniques, as discussed in \cite{Huterer} for Type Ia supernova data and in \cite{Amendola} for fine-structure constant measurements. This method has been numerically implemented in a code originally developed by Nelson Nunes for \cite{Amendola}, subsequently used in \cite{Leite}, and now modified and extended for the present work. We can now proceed to discussing our assumptions regarding the fiducial models and simulated datasets.

\section{\label{assumptions}Models and data}

We will assume a flat universe, and further simplify the analysis by fixing $\Omega_m=0.3$. This is a standard procedure, followed in many previous works including \cite{Huterer,Amendola,Leite}. This specific choice of $\Omega_m$ has a negligible effect on the main results of our analysis. In any case, we note that the purpose of these PCA techniques is to characterize gains in sensitivity, rather than provide hard numbers.

We will consider two fiducial models for $w(z)$, already used in \cite{Amendola,Leite}
\begin{equation}
w_{c}(z)=-0.9,
\end{equation}
\begin{equation}
w_{s}(z)=-0.5+0.5 \tanh \left(z-1.5 \right)\,.
\end{equation}
At a phenomenological level, these describe two opposite and qualitatively different scenarios: an equation of state that remains close to a cosmological constant throughout the probed redshift range, and one that evolves towards a matter-like behavior by the highest redshifts probed. In what follows we will refer to these cases as the \textit{Constant} and \textit{Step}. For each fiducial model we choose a prior for the coupling such that it leads to a few parts-per-million variation of $\alpha$ at redshift $z\sim4$, consistently with \cite{Dipole}. 

For Type Ia supernovas we will consider the following datasets:
\begin{itemize}
\item A low-redshift sample, henceforth denoted \textbf{LOW}, of 3000 supernovas uniformly distributed in the redshift range $0<z<1.7$, with an uncertainty on the magnitude of $\sigma_m = 0.11$. These numbers are typical of 'SNAP-like' future supernova datasets and were also used in \cite{Huterer} (thus providing a useful point of comparison).
\item An intermediate redshift sample, henceforth denoted \textbf{MID}, of 1700 supernovas uniformly distributed in the redshift range $0.75<z<1.5$ and with the same $\sigma_m$ as before. This is representative of recent proposals such as DESIRE \cite{Desire}, and forecasts with a similar sample were already discussed (in different contexts) in \cite{Tasos,Erminia}.
\item A high-redshift sample of supernovas identified by JWST NIRcam imaging \cite{JWST} and then characterized by extremely large telescopes on the ground such as the  E-ELT and the TMT. Based on their respective Phase A studies we assume a sample of 50 supernovas in the range $1<z<5$ for the E-ELT \cite{HARMONI} and a sample of 250 supernovas in the range $1<z<3$ for the TMT \cite{TMT}. These will be denoted \textbf{ELT} and \textbf{TMT} respectively, and they will provide a useful proxy for studying the importance of the redshift lever arm versus the size of the sample. The redshift distribution of these supernovas is not easy to extrapolate, since even the most detailed current studies such as those of the SNLS \cite{SNLS} only reach $z\sim1$, but in the absence of more detailed information we again assume a uniform distribution in the respective redshift ranges and the same $\sigma_m$ as before. Forecasts with a similar sample were already discussed (in different contexts) in \cite{Tasos}.
\end{itemize}

For the fine-structure constant measurements we will focus on the ESPRESSO spectrograph for the VLT \cite{ESPRESSO} and its successor ELT-HIRES for the E-ELT \cite{HIRES}; in the tables that follow we will denote them \textbf{ESP} and \textbf{HRS} respectively. Both spectrographs include tests of the stability of fundamental constants among their key science and design drivers. A more detailed discussion and roadmap for these tests can be found in \cite{GRG}. Specifically we will consider the following two scenarios
\begin{itemize}
\item A \textbf{Baseline} scenario, in which we will assume measurements in 30 systems with uncertainty $\sigma_{\Delta\alpha/\alpha} = 6\times10^{-7}$ for ESPRESSO and 100 systems with $\sigma_{\Delta\alpha/\alpha} = 1\times10^{-7}$ for ELT-HIRES, uniformly distributed in the redshift range $0.5<z<4$. This is meant to represent what we can confidently expect to achieve from each spectrograph (e.g., from Guaranteed Time Observations), given the their expected sensitivity, and it will therefore provide the basis for most of our discussion.
\item An \textbf{Ideal} scenario, in which we will assume 100 systems with $\sigma_{\Delta\alpha/\alpha} = 2\times10^{-7}$ for ESPRESSO and 150 systems with $\sigma_{\Delta\alpha/\alpha} = 3\times10^{-8}$ for ELT-HIRES. This is optimistic both in the uncertainty of individual measurements and in the number of measurements. Although several hundred absorbers are already known where these measurements can be carried out, the sources are quite faint and---as can be extrapolated from current VLT data \cite{Leite}---putting together such a dataset would at the very least require a very long time and almost certainly a dedicated program. Our goal in considering this ideal case is to obtain an indication for the dependence of our results on the uncertainty and number of the measurements.
\end{itemize}
We note that these are two of the cases already studied in \cite{Amendola}, whose analysis we will extend.

For our PCA analysis we will in general assume 30 redshift bins in the range $0\le z\le4$. We note that this redshift range is conservative for $\alpha$ measurements, since measurements beyond $z=4$ already exist, and the infrared sensitivity of ELT-HIRES may push this limit even further \cite{HIRES}. In the case where ELT supernovas are used, the last bin is extended until $z=5$.

For comparison purposes we will also briefly consider a case with only 20 bins. This serves to provide an illustration of the effects of redshift resolution on the reconstruction. Using too few bins is likely to erase useful information (especially if the behavior of the dark energy equation of state is non-trivial), while using too many will lead to very little observational information in some (or all) bins, with runs the risk of misinterpreting noise as non-trivial information. To a large extent the choice of the optimal number of bins will depend on the available datasets themselves; while we won't address this issue explicitly here, we nevertheless quote the results with the two choices of bins to provide the reader with an illustration of the importance of an adequate choice.

In order to quantify gains in sensitivity we will use two different diagnostics. The simplest one is the number of PCA modes with uncertainties below $\sigma_{PCA}=0.3$; although this choice of threshold is somewhat arbitrary, it has already been used in \cite{Crittenden} and in our previous analysis \cite{Amendola}.

Tables \ref{modes1} and \ref{modes2} display these numbers, respectively for the Constant and Step fiducial models, assuming the Baseline scenario for $\alpha$ measurements, and for various combinations of supernova datasets. Considering supernova data only, the MID sample adds one mode, and the further inclusion of ELT supernovas may add an additional one. When combining supernova and $\alpha$ measurements, ESPRESSO may add up to two modes while ELT-HIRES adds many more. For the combined datasets, whether the TMT sample (more supernovas at lower redshifts) or the E-ELT one (fewer supernovas at higher median redshift) is the more informative one is model-dependent. The reason why more modes are well characterized in the Constant than in the Step case is that, with our choice of high-redshift normalization for the $\alpha$ variations, a uniform distribution of the measurements in redshift turns out to be an optimal observational strategy for the Constant case, but is far from optimal for the Step case; this is further discussed in \cite{Leite}. Here we have chosen to keep the assumption of uniform redshift sampling precisely to highlight this model-dependence.

A somewhat more informative diagnostic is 'figure of merit' defined as the inverse of the product of the uncertainties of the two best determined modes, $FoM=1/(\sigma_1\sigma_2)$ \cite{FOM}. We will adopt this for a more thorough exploration of these scenarios in the following section.


\begin{table}[htbp]
\small
  \centering
  \caption{\label{modes1}Number of PCA modes with uncertainties below $\sigma_{PCA}=0.3$, assuming the 'Constant' fiducial model, the 'Baseline' scenario for $\alpha$ measurements, and 30 redshift bins.}
		\begin{tabular}{|l|c|c|c|}		
		\hline
          & Sne only & Sne + ESP & Sn + HRS \\
    \hline
    LOW  & 3     & 5     & 17  \\
    \hline
    LOW + MID & 4     & 5     & 18  \\
    \hline
    LOW + ELT & 4     & 5     & 16  \\
    \hline
    LOW + MID + ELT & 4     & 6     & 16  \\
    \hline
    LOW + TMT & 4     & 5     & 17  \\
    \hline
    LOW + MID + TMT & 4     & 5     & 18  \\
    \hline
    \end{tabular}%
\end{table}%

\begin{table}[htbp]
\small
  \centering
  \caption{\label{modes2}Number of PCA modes with uncertainties below $\sigma_{PCA}=0.3$, assuming the 'Step' fiducial model, the 'Baseline' scenario for $\alpha$ measurements, and 30 redshift bins.}
		\begin{tabular}{|l|c|c|c|}		
		\hline
           & Sne only & Sne + ESP & Sne + HRS \\
   \hline
    LOW  & 3     & 4     & 9  \\
    \hline
    LOW + MID & 4     & 4     & 10  \\
    \hline
    LOW + ELT & 4     & 4     & 10  \\
    \hline
    LOW + MID + ELT & 5     & 5     & 10  \\
    \hline
    LOW + TMT & 4     & 4     & 9  \\
    \hline
    LOW + MID + TMT & 5     & 5     & 11  \\
    \hline
    \end{tabular}%
\end{table}%



\section{\label{results}Comparing Figures of Merit}

The figures of merit for the baseline case of $\alpha$ measurements are shown in Tables \ref{fombase1} and \ref{fombase2}, respectively for the Constant and Step fiducial models. In both cases we compare the results obtained with 20 or 30 bins. Table \ref{fomideal} shows the results for the Ideal $\alpha$ datasets and 30 redshift bins, comparing the results for the two fiducial models.

We note that the gains in sensitivity to the dark energy equation of state due to ESPRESSO measurements are relatively modest in the Baseline case, but significant (up to about a factor of 2) in the Ideal case. ELT-HIRES, on the other hand, will lead to dramatic improvements (sometimes more than a factor of 50). These results are consistent with the findings of \cite{Leite}, whose analysis used different criteria.

It's also noteworthy that judging by this figure of merit diagnostic the impact of the E-ELT supernovas is always greater than that of the TMT supernovas. This is the case whether one is using supernova data only or a combination of supernovas and $\alpha$ measurements. Note that for the case of supernovas only, the 50 ELT supernovas (uniformly distributed in the range $1<z<5$) would not only be more constraining than the 250 TMT supernovas (in the range $1<z<3$) but also more constraining than the 1700 MID supernovas (in the range $0.75<z<1.5$).

We caution the reader that there are obvious caveats to this comparison. Firstly, finding very high redshift supernovas will be difficult, and current estimates of expected rates are at best uncertain. Moreover, the E-ELT or TMT time required to characterize them will certainly be costly, and we are taking for granted a temporal overlap between JWST and the relevant E-ELT and TMT instruments. Nevertheless, the results of this comparison do highlight the importance of the redshift lever arm in characterizing dynamical dark energy.

Finally, we also briefly studied how these results will be affected if the intrinsic dispersion of the supernova magnitudes is $\sigma_m=0.15$ (instead of the $\sigma_m=0.11$ that is assumed in the rest of the paper.) In order to quantify this we have repeated the calculation of the figures of merit listed in Table \ref{fombase2} with this degraded dispersion, and the revised results are listed in Table \ref{fom15}. We see that the absolute numbers change significantly (by up to a factor of 2 with supernova data alone, by 25 to 35 percent for ELT-HIRES), but the relative variations between the various cases maintain the previously discussed behavior. Nevertheless it's noteworthy that the figures of merit decrease more for the case of 30 bins, as was to be expected.


\begin{table}[htbp]
\small
  \centering
  \caption{\label{fombase1}Figures of merit for the dark energy equation of state, assuming the 'Constant' fiducial model and the 'Baseline' scenario for $\alpha$ measurements, and 30 redshift bins. For each pair of entries the top and bottom lines respectively assume 20 and 30 redshift bins.}
		\begin{tabular}{|l|c|c|c|}		
		\hline
          & Sne only & Sne + ESP & Sne + HRS \\
		\hline
    LOW (20)  & 539  & 546  & 5215 \\
    LOW (30) & 409  & 412  & 3574  \\
		\hline
    LOW + MID (20) & 1090  & 1096  & 5331 \\
    LOW + MID (30) & 839  & 843  & 3655  \\
		\hline
    LOW + ELT (20) & 1194  & 1215  & 8055 \\
    LOW + ELT (30) & 881  & 888  & 4947  \\
		\hline
    LOW + MID + ELT (20) & 2371  & 2392  & 8493 \\
    LOW + MID + ELT (30) & 1973  & 1980  & 5286  \\
		\hline
    LOW + TMT (20) & 808  & 814  & 5302 \\
    LOW + TMT (30) & 631  & 634  & 3642  \\
		\hline
    LOW + MID + TMT (20) & 1581  & 1586  & 5520 \\
    LOW + MID + TMT (30) & 1253  & 1256  & 3814  \\
 		\hline
    \end{tabular}%
\end{table}%

\begin{table}[htbp]
\small
  \centering
  \caption{\label{fombase2}Figures of merit for the dark energy equation of state, assuming the 'Step' fiducial model and the 'Baseline' scenario for $\alpha$ measurements, and 30 redshift bins. For each pair of entries the top and bottom lines respectively assume 20 and 30 redshift bins.}
  \begin{tabular}{|l|c|c|c|}
		\hline
          & Sne only & Sne + ESP & Sne + HRS \\
		\hline
    LOW (20)  & 536  & 541  & 1358\\
    LOW (30)  & 404  & 407  & 982  \\
		\hline
    LOW + MID (20) & 1084  & 1089  & 2003 \\
    LOW + MID (30) & 831  & 834  & 1462  \\
		\hline
    LOW + ELT (20) & 1225  & 1243  & 3206 \\
    LOW + ELT (30) & 881  & 885  & 1738  \\
		\hline
    LOW + MID + ELT (20) & 2432  & 2450  & 2561 \\
    LOW + MID + ELT (30) & 2175  & 2176  & 2356  \\
		\hline
    LOW + TMT (20) & 821  & 824  & 1453 \\
    LOW + TMT (30) & 634  & 636  & 1055  \\
		\hline
    LOW + MID + TMT (20) & 1605  & 1608  & 2209 \\
    LOW + MID + TMT (30)  & 1260  & 1262  & 1636  \\
		\hline
    \end{tabular}%
\end{table}%

\begin{table}[htbp]
\small
  \centering
  \caption{\label{fomideal}Figures of merit for the dark energy equation of state, assuming the 'Ideal' scenario for $\alpha$ measurements and 30 redshift bins. For each pair of entries the top and bottom lines respectively correspond to the Constant and Step fiducial models.}
  \begin{tabular}{|l|c|c|c|}
                \hline
          & Sne only & Sne + ESP & Sne + HRS \\
                \hline
    LOW (c)  & 409  & 996  & 58684  \\
    LOW (s)  & 404  & 554  & 11228  \\
                \hline
    LOW + MID (c) & 839  & 1352  & 58737  \\
    LOW + MID (s) & 831  & 955  & 11295  \\
                \hline
    LOW + ELT (c) & 881  & 1515  & 79431  \\
    LOW + ELT (s) & 881  & 1064  & 18176  \\
                \hline
    LOW + MID + ELT (c) & 1973  & 2357  & 79639  \\
    LOW + MID + ELT (s) & 1971  & 2133  & 18652  \\
                \hline
    LOW + TMT (c) & 631  & 1089  & 58740  \\
    LOW + TMT (s) & 634  & 712  & 11335  \\
                \hline
    LOW + MID + TMT (c) & 1253  & 1443  & 58846  \\
    LOW + MID + TMT (s) & 1260  & 1328  & 11514  \\
                \hline

    \end{tabular}%
\end{table}%

\begin{table}[htbp]
\small
  \centering
  \caption{\label{fom15}Same as Table \protect\ref{fombase2}, but now assuming an intrinsic dispersion of the supernova magnitudes of $\sigma_m=0.15$ rather than $\sigma_m=0.11$}
		\begin{tabular}{|l|c|c|c|}		
		\hline
          & Sne only & Sne + ESP & Sne + HRS \\
		\hline
    LOW (20)  & 303  & 309  & 1017 \\
    LOW (30) & 229  & 232  & 730  \\
		\hline
    LOW + MID (20) & 613  & 618  & 1459 \\
    LOW + MID (30) & 470  & 473  & 1065  \\
		\hline
    LOW + ELT (20) & 768  & 769  & 1279 \\
    LOW + ELT (30) & 570  & 571  & 936  \\
		\hline
    LOW + MID + ELT (20) & 1605  & 1606  & 1933 \\
    LOW + MID + ELT (30) & 1230  & 1231  & 1450  \\
		\hline
    LOW + TMT (20) & 464  & 468  & 1125 \\
    LOW + TMT (30) & 359  & 361  & 807  \\
		\hline
    LOW + MID + TMT (20) & 908  & 911  & 1541 \\
    LOW + MID + TMT (30) & 713  & 715  & 1130  \\
 		\hline
    \end{tabular}%
\end{table}%


One can reconstruct $w(z)$ by keeping only the most accurately determined modes, as was first done, for supernovas only, in \cite{Huterer}. (See \cite{Crittenden} for an alternative approach.) To do this, we need to decide how many components to keep. The canonical choice for the optimal value of modes $M$ to be kept corresponds to the value that minimizes the risk, defined as \cite{Huterer}
\begin{equation}
R = Bias^2 + Var= \sum_{i=1}^N\left( \tilde w(z_i) - w_\star(z_i) \right)^2+ \sum_{i=1}^N \sum_{j=1}^M \sigma^2(\alpha_j) e_j(z_i)\,,
\end{equation} 
where the notation $\tilde w$ means that the sum in Eq. (\ref{recw}) runs from 1 to $M$ and $w_\star$ denotes the relevant fiducial model. The bias measures how much the reconstructed equation of state differs from the true one by neglecting the noisy modes, and typically decreases as we increase $M$. Conversely the variance of $w(z)$ increases as we increase $M$, since we will be including modes that are less accurately determined.

An alternative, previously used in a similar context in \cite{Amendola}, is to choose the largest value for which the error is below unity, or equivalently, the RMS fluctuations of the equation of state parameter in such a mode are
\begin{equation}
\langle (1+w(z))^2 \rangle = \sigma_i^2 \lesssim 1\,
\end{equation}
and then normalize the errors as suggested in \cite{FOM}, such that $\sigma^2 = 1$ for the worse determined mode and
\begin{equation}
\sigma^2(\alpha_i) \rightarrow \sigma_n^2(\alpha_i) =
\frac{\sigma^2(\alpha_i)}{1+\sigma^2(\alpha_i)} .
\end{equation}

As has been discussed in \cite{Amendola}, one expects that the error normalization method will generically select more modes than risk minimization. This will therefore lead to a more accurate reconstruction (ie, closer to the correct fiducial model), though correspondingly also one with larger error bars. The normalization method is therefore the more conservative one, while the risk method is more aggressive. We will further quantify this below.

Importantly, because we truncate the above sum (neglecting the poorly determined modes with high amplitudes at larger redshift) the reconstructed equation of state necessarily tends to zero for sufficiently large redshift \cite{Huterer}. This unavoidable feature of the PCA truncation method can be confused with a real increase in the equation of state at high redshift. This is another reason for wanting to extend the redshift range and the sensitivity of the measurements: they will lead to more reliable reconstructions at higher redshifts.

\begin{figure*}
\begin{center}
\includegraphics[width=6in]{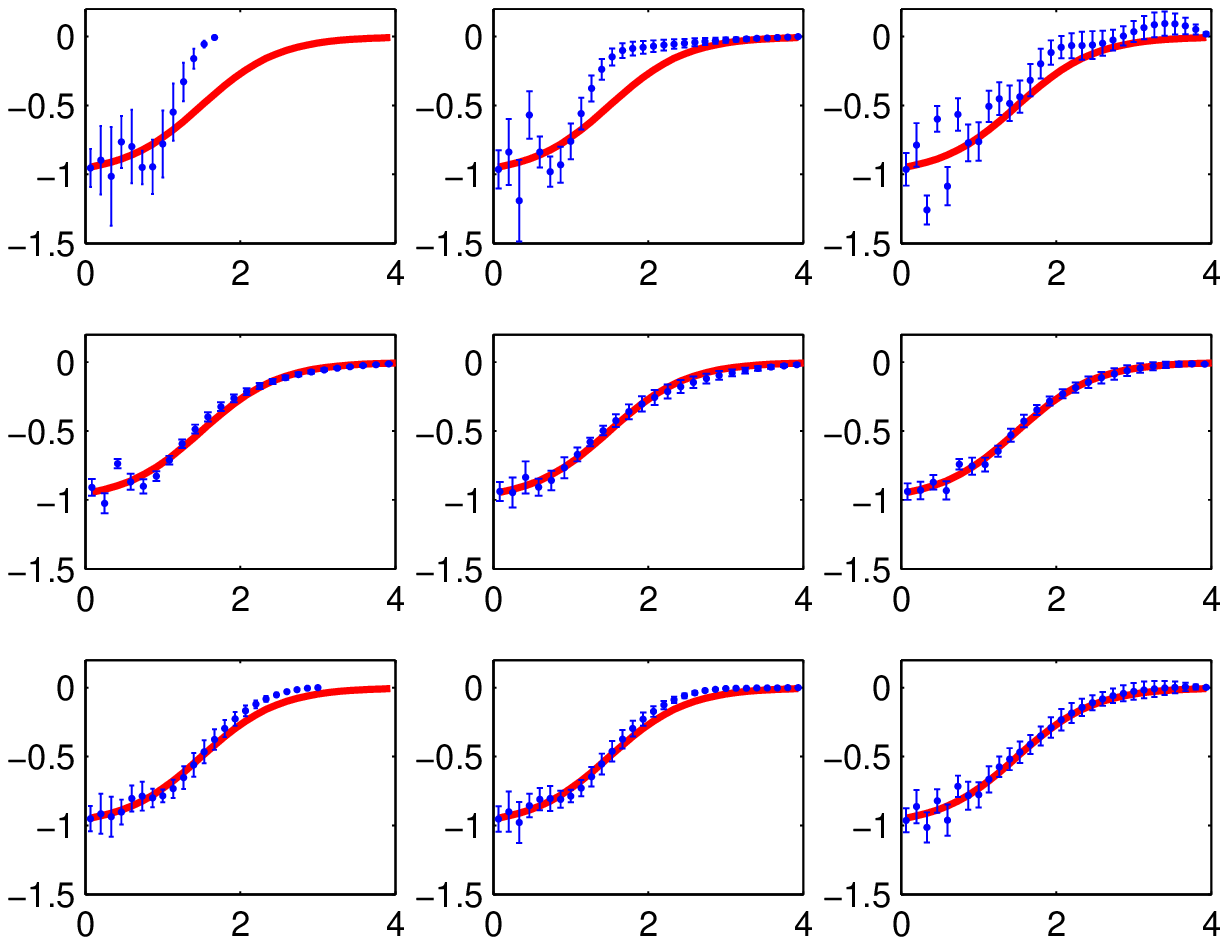}
\includegraphics[width=6in]{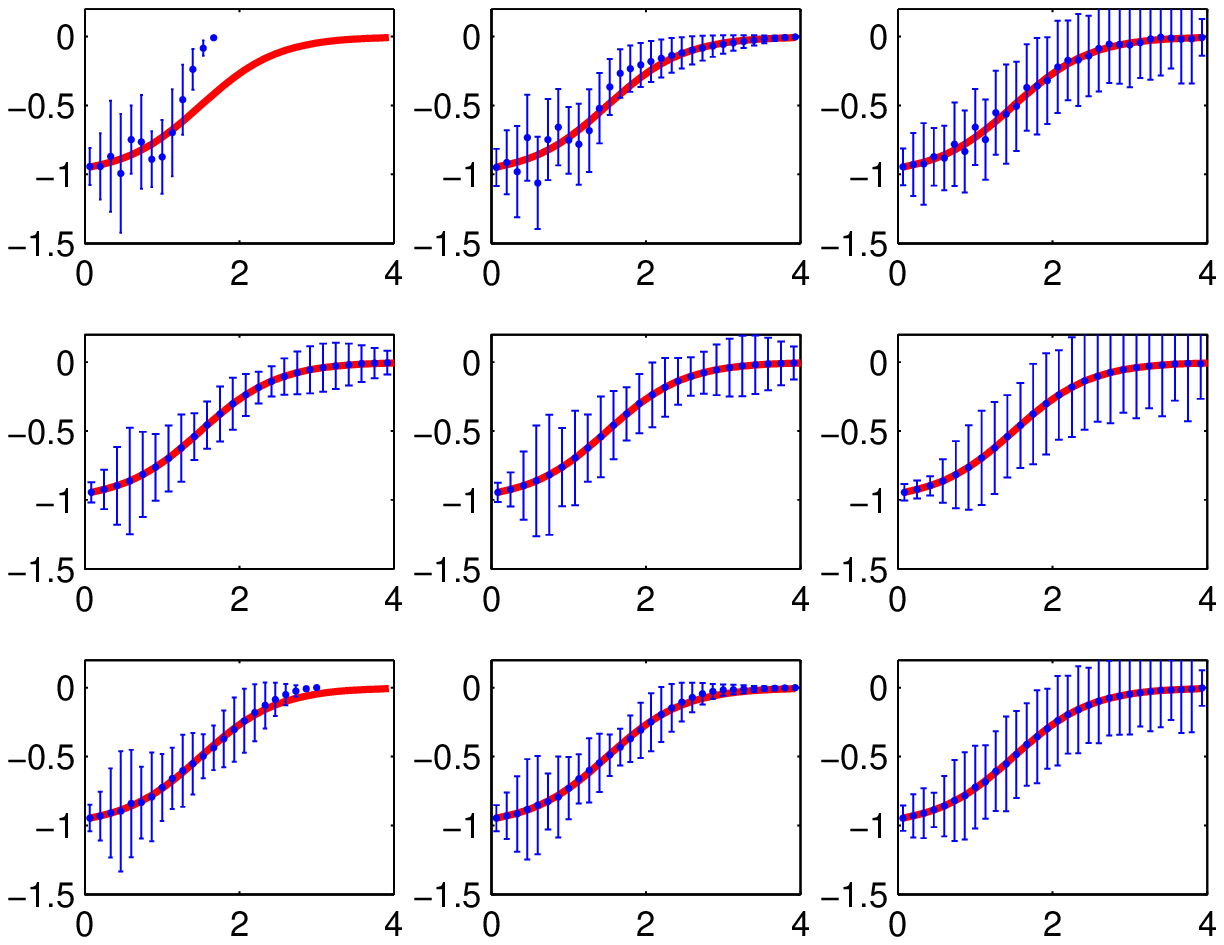}
\end{center}
\caption{\label{fig1}Examples of dark energy equation of state reconstructions for the {\it Step} fiducial model, using the risk minimization and normalization methods (top and bottom set of plots, respectively. All panels show the equation of state $w(z)$ plotted as a function of redshift. In each set of plots the left panels correspond to supernova data only, the middle ones to the combined supernova and ESPRESSO data, and the right ones to the combined supernova and ELT-HIRES data, while the three lines correspond to LOW+MID, LOW+MID+ELT and LOW+MID+TMT. In all cases we assumed the Baseline scenario for $\alpha$ measurements.}
\end{figure*}
\begin{figure*}
\begin{center}
\includegraphics[width=6in]{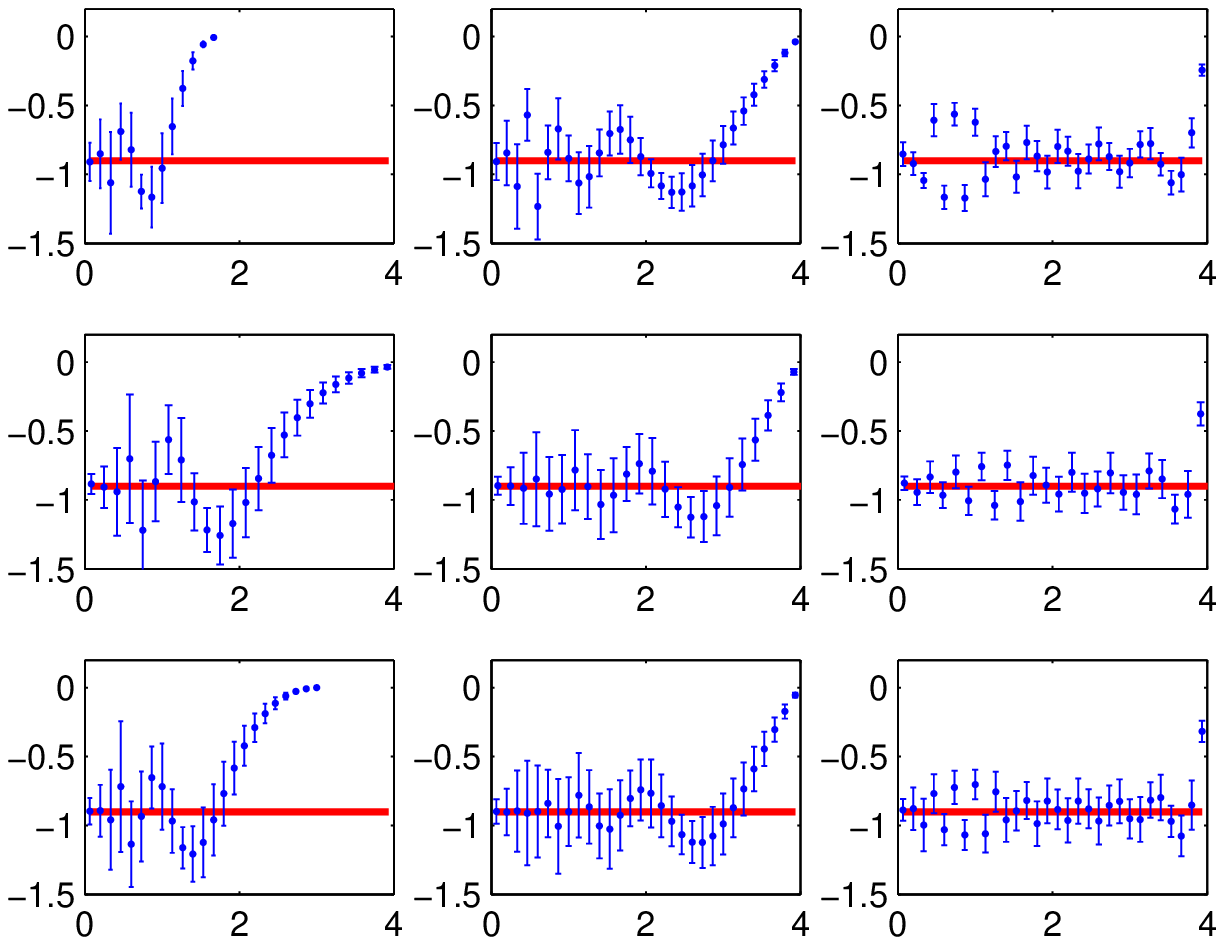}
\includegraphics[width=6in]{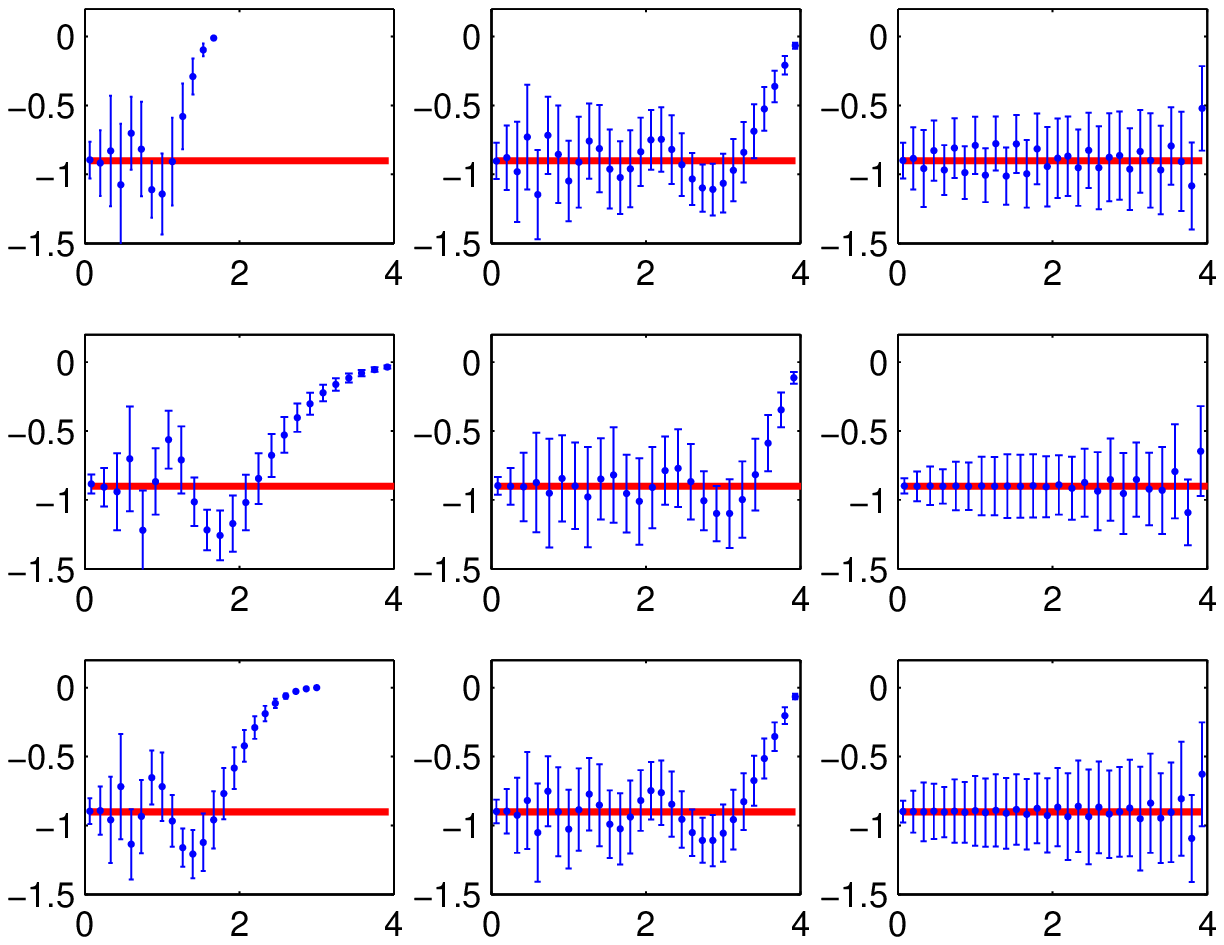}
\end{center}
\caption{\label{fig2}Examples of dark energy equation of state reconstructions for the {\it Constant} fiducial model, using the risk minimization and normalization methods (top and bottom set of plots, respectively. All panels show the equation of state $w(z)$ plotted as a function of redshift. In each set of plots the left panels correspond to supernova data only, the middle ones to the combined supernova and ESPRESSO data, and the right ones to  the combined supernova and ELT-HIRES data, while the three lines correspond to LOW+MID, LOW+MID+ELT and LOW+MID+TMT. In all cases we assumed the Baseline scenario for $\alpha$ measurements.}
\end{figure*}
\begin{figure*}
\begin{center}
\includegraphics[width=6in]{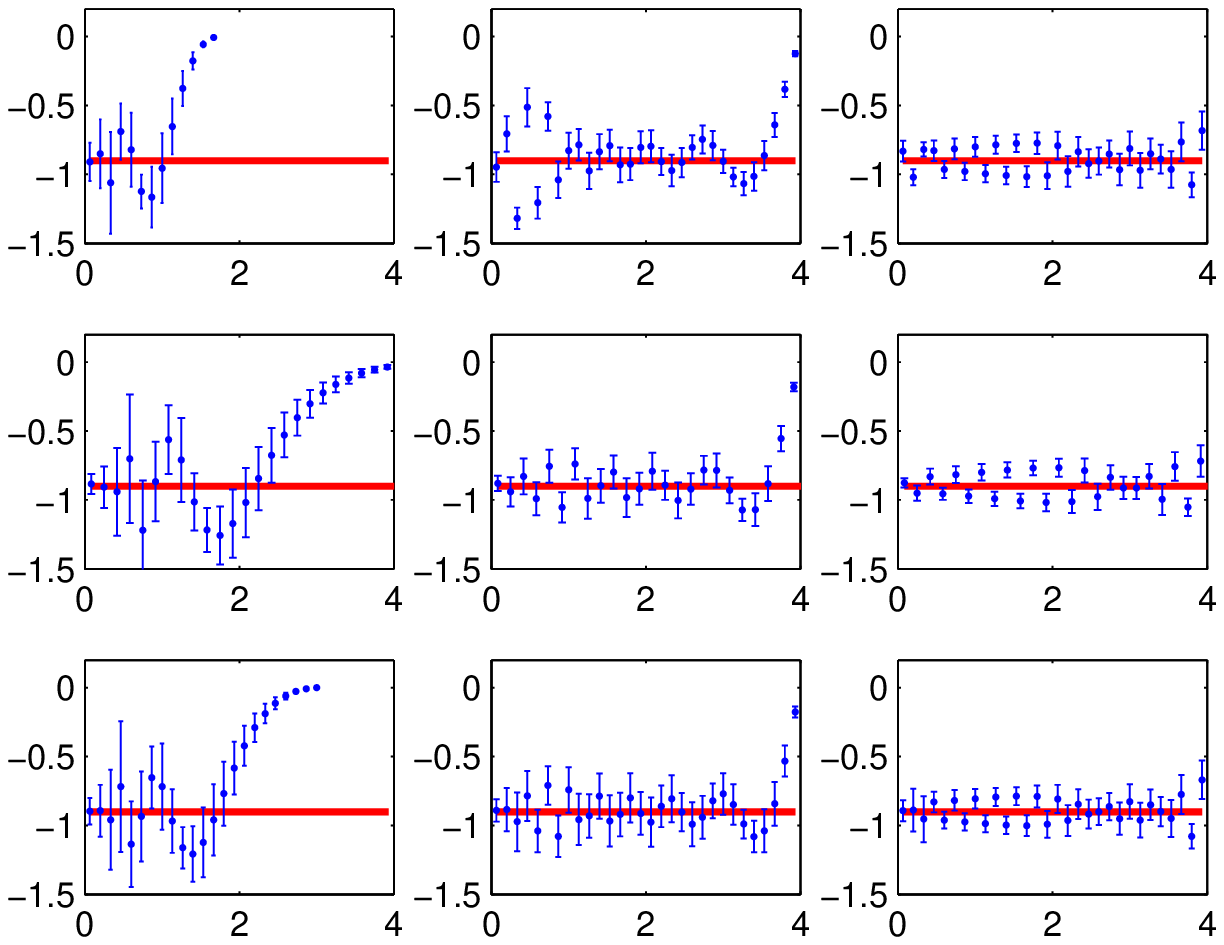}
\includegraphics[width=6in]{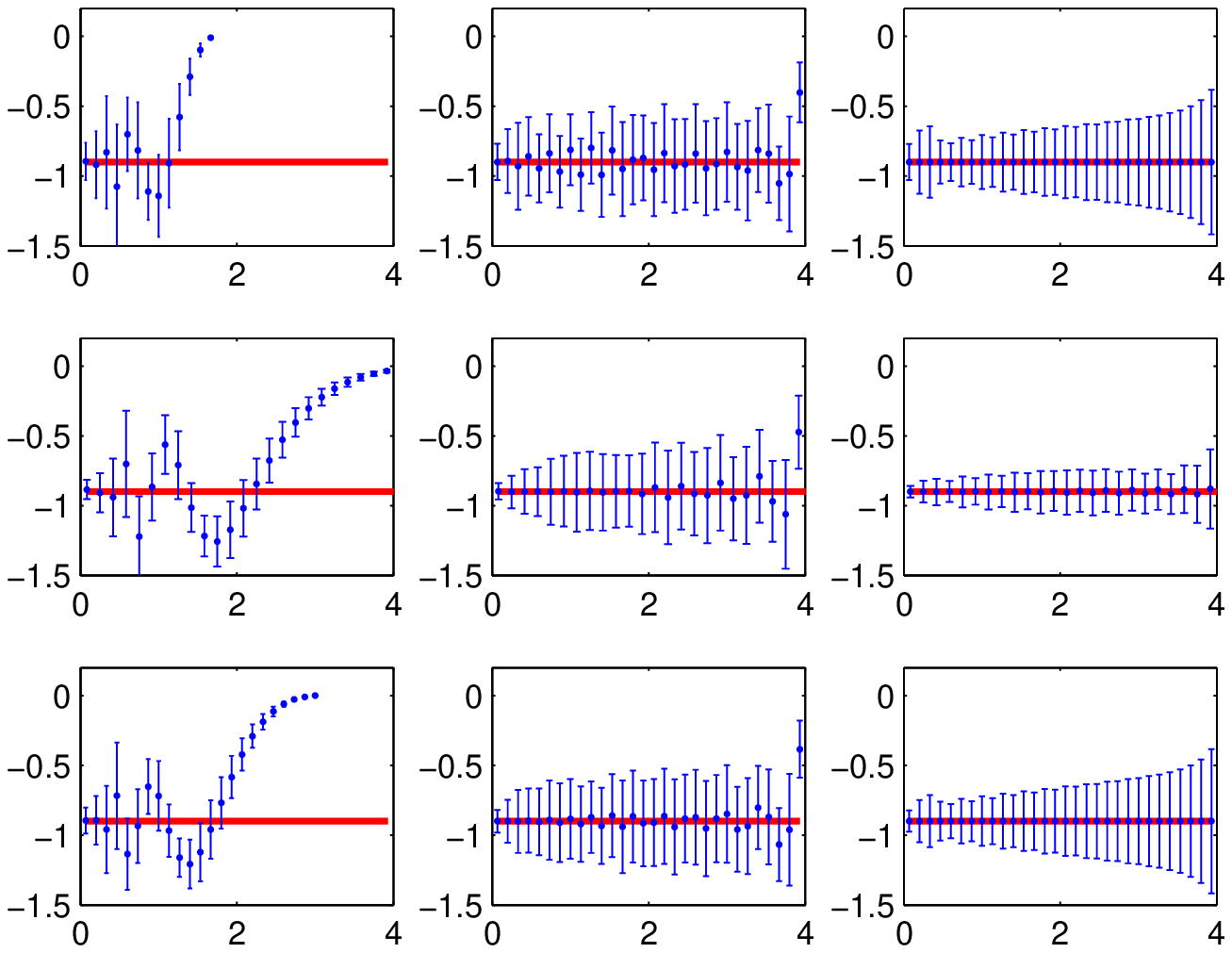}
\end{center}
\caption{\label{fig3}Further examples of dark energy equation of state reconstructions for the {\it Constant} fiducial model. Parameter choices are identical to those of Fig. \protect\ref{fig2}, except that we now assumed the Ideal scenario for $\alpha$ measurements.}
\end{figure*}

Figure \ref{fig1} shows example of reconstructions, for various choices of datasets, for the {\it Step} model, using either method to truncate the series; these correspond, respectively, to the top and bottom set of 9 panels. In both cases we assumed the Baseline scenario for $\alpha$ measurements. Note that in this and subsequent plots we always plot the reconstructed equation of state until redshift $z=4$, even though for the E-ELT supernova case there will be measurements until $z=5$. (In a given figure the same mocks have been used for all panels, but different mocks were used for different figures; this choice of mocks does not significantly affect our results.)

Naturally the step model provides a best-case scenario, since the correct fiducial equation does approach zero at high redshifts. This comparison makes it clear that the risk minimization method leads to a reconstruction with nominally very small error bars, but also one that may be somewhat biased. It is clear from the plots that this bias can be decreased either by extending the redshift lever arm (compare the left-most column, for reconstructions with supernovas only) or by increasing their sensitivity (compare the middle and right-most set of plots, which respectively include ESPRESSO and ELT-HIRES $\alpha$ measurements). On the other hand the normalization method in this case yields very conservative error bars and an almost perfect reconstruction, as long as one has a sufficient amount of data deep in the matter era.

Conversely the case of the {\it Constant} equation of state provides a worst-case scenario, as shown in Fig. \ref{fig2}. Here the reconstruction will necessarily be biased at sufficiently high redshifts. In other words, the reliability of the reconstruction at high redshift will be model-dependent. This much is of course expected, but these plots again make it clear that the reliability can always be increased (all else being equal) by extending the redshift range where data is available and improving their sensitivity---compare the reconstructions using ESPRESSO and ELT-HIRES in this figure. Since one does not a priori know the high-redshift behavior of the equation of state, going as deep in redshift as possible is a mandatory aspect of any observational strategy optimization. Tests of the stability of the fine-structure constant can therefore play a key role in this endeavor. It remains true in this case that the normalization method to truncate the series leads to a more robust reconstruction.

In order to further quantify the impact of some of our assumptions on the above results, Fig. \ref{fig3} shows the results of two other reconstructions for the constant fiducial model case, where we now assumed the Ideal scenario for $\alpha$ measurements. This comparison (which obviously will only affect the cases with $\alpha$ measurements) clearly shows that with the increased sensitivity and number of measurements of the Ideal scenario the reliability of the reconstruction is significantly improved, both in terms of error bars in the various redshift bins and in terms of the maximum value of the redshift where the reconstruction is not significantly biased. The panels corresponding to the {\it Constant} model are particularly illuminating in illustrating the relative contributions of the supernova and $\alpha$ datasets: the difference between the E-ELT and TMT supernovas are clearly visible, and indeed even enhanced in this case by the presence of similarly sensitive E-ELT $\alpha$ measurements.


\section{\label{concl}Conclusions}

We have used previously available PCA-based forecast techniques to quantify the gains in sensitivity expected from constraints on the behavior of dark energy enable by forthcoming ground and space-based astronomical facilities. Specifically we have focused on the reconstruction of the dark energy equation of state, using both future space-based supernova surveys in combination with high-resolution spectroscopic measurements of the fine-structure constant expected from ESPRESSO (whose commissioning at the VLT will be in late 2016) and the high-resolution ultra-stable spectrograph planned for the E-ELT (and currently known as ELT-HIRES).

Our results quantitatively confirm that the combination of these two types of measurements, which probe different (but overlapping) redshift ranges leads to a more complete---and robust---mapping of the evolution of the equation of state, and that a detailed reconstruction between redshift zero and four is within the reach of forthcoming facilities. The combination of the two datasets leads to figure of merit improvements that are typically a factor of a few, and more than 50 in ideal circumstances. We also provided a further comparison (for our fiducial models) of the two PCA truncation criteria. We would argue that the more conservative normalization should be the default method for truncation, unless there is some prior evidence suggesting that the dark equation of state is diverging from -1 at high redshifts (in other words, broadly speaking, that one is dealing with a freezing model rather than a thawing one).

Finally, we should point out an additional possibility not directly addressed in our work. A comparison of the two reconstructions (separately using supernovas and $\alpha$ measurements) can provide a consistency test, and more specifically it will be a test for the assumption that the same dynamical degree of freedom is responsible for the dark energy and the $\alpha$ variation---in other words, that one is dealing with a Class I model, in the terminology of \cite{GRG}. Assuming this is the case, then the coupling between the scalar field and the electromagnetic sector should be describable (at least to first approximation) by Eq.~\ref{coupling}, and one can also infer the posterior likelihood for the coupling parameter $\zeta$. A detailed study of this interesting possibility is left for forthcoming work.

\begin{acknowledgments} 
We are grateful to Luca Amendola, Nelson Nunes and Pedro Pedrosa for many interesting discussions and suggestions in the early stages of this project. This work was done in the context of grant PTDC/FIS/111725/2009 from FCT (Portugal). A.C.L. is supported by the Gulbenkian Foundation (Portugal) through {\it Programa de Est\'{\i}mulo \`a Investiga\c c\~ao 2014}, grant number 2148613525. C.J.M. is also supported by an FCT Research Professorship, contract reference IF/00064/2012, funded by FCT/MCTES (Portugal) and POPH/FSE (EC).
\end{acknowledgments}

\bibliography{pca}

\end{document}